# Ultrasmall In(Ga)As Quantum Dots Grown on Strained $Ga_xIn_{1-x}P$ Layers

T. Raz,* D. Ritter and G. Bahir

*Department of Electrical Engineering, Technion - Israel Institute of Technology, Technion City, Haifa 32000, Israel*

D. Gershoni

*The Physics Department and The Solid State Institute, Technion - Israel Institute of Technology, Technion City, Haifa 32000, Israel*

**ABSTRACT**

InAs self-assembled quantum dots were grown on strained layers of $Ga_xIn_{1-x}P$ ($0 < x < 0.3$) on InP substrates. We show that the quantum dots have narrow vertical dimensions, ranging between 2 to 10 monolayers only. The dot layer photoluminescence spectrum is composed of distinct spectral peaks, resulting from the discrete distribution of the dot heights. The dot height distribution depends on the total amount of InAs deposited and on the Ga content of the strained $Ga_xIn_{1-x}P$ layer underneath. Our experimental results are corroborated by an 8 band k·P model calculations. In particular, we identify and quantify the diffusion of Ga from the $Ga_xIn_{1-x}P$ layer into the quantum dots.

* Electronic mail: talraz@tx.technion.ac.il

The growth of self-assembled quantum dots (SAQDs) due to the strain generated by the lattice mismatch between the dot layer material and the substrate material is well-established.[1] The most thoroughly investigated material system is that of InAs SAQDs grown on GaAs surfaces, where the lattice mismatch is 7.2%. The formation of InAs SAQDs on InP substrates, where the lattice mismatch amounts to 3.2% only, is studied to a lesser extent.[2-5] Typically, the height distribution of an ensemble of SAQDs in both systems is in the range of 10 – 30 monolayers (MLs).[2] Thus, a single SAQD contains about $10^4 – 10^5$ atoms. Under certain growth conditions, however, InAs SAQDs having heights in the range of 4 to 8 monolayers can be grown on InP.[4,6] In this letter, we show that by introducing a thin tensile strained layer of $Ga_xIn_{1-x}P$ between the InP substrate and the InAs SAQD layer, the height of the SAQDs may be made as short as two monolayers only.

The samples were grown by a compact metalorganic molecular beam epitaxy (MOMBE) system[7] at 510ºC. Trimethylindium, triethylgalium, arsine, and phosphine served as group III and V sources, respectively. A 200 nm thick InP buffer layer was grown, followed by a $Ga_xIn_{1-x}P$ layer, a single layer of InAs QDs, and a 50 nm thick InP cap layer. The growth rates of the InAs, InP and the $Ga_xIn_{1-x}P$ are 0.12, 1.5 and 1.85 ML/s, and the V/III ratios are 30, 1.8 and 2.55, respectively. The growth was interrupted for 40 sec prior to and after the SAQDs layer growth. The thickness of the $Ga_xIn_{1-x}P$ layers was kept well below the critical thickness for strain relaxation,[8,9] to ensure that they are fully strained. Uncapped SAQD samples were grown under the same conditions, in order to measure their density by an atomic force microscope.

Two series of samples were studied. The first one consisted of InAs with a nominal thickness ranging from 1.2 to 4.2 MLs grown on a 10 nm thick $Ga_{0.19}In_{0.81}P$ layer. The second one consisted of nominally 2.4 ML InAs grown on a 10 nm thick layer of $Ga_xIn_{1-x}P$ with various x values ranging from 0 to 0.3. The sample with x = 0.3 had only 5 nm thick $Ga_{0.3}In_{0.7}P$ layer. The density of the QDs grown on $Ga_{0.19}In_{0.81}P$ (first series) increased gradually with increasing InAs thickness, from $3\times10^8$ $cm^{-2}$ (1.2 ML) to $7\times10^9$ $cm^{-2}$ (4.2 ML). With increasing Ga content in the underlying $Ga_xIn_{1-x}P$ layer, the dot density increased only slightly.

The 77K photoluminescence (PL) spectra of the first series of samples are shown in Fig. 1. The spectra are vertically displaced for clarity. The uppermost curve describes the PL spectrum from the sample where the InAs nominal thickness was 1.2 ML only. There is only one peak in this spectrum, centered at 1.265 eV. We identify this spectral line as due to electron–hole (e-h) recombination within the wetting layer (WL). The spectra below are from samples with successively thicker InAs layer. The PL from the WL drops rapidly in these spectra, while additional PL lines appear at lower energies. We identify these PL lines as due to e-h recombination within SAQDs of monolayer step different heights. Thus, the closest peak to the WL PL line is attributed to PL from 2 monolayers high SAQDs, the next lower energy peak is attributed to 3 monolayers high SAQDs, and so on. We fitted a multi-Gaussian model to the experimentally measured spectra, as shown in the lower most-spectrum in Fig. 1, in order to determine the various PL line energies and distributions.

We note here that distinct peaks in the PL spectrum of SAQDs are usually attributed to emission due to recombination of e-h pairs from higher energy states of the SAQDs. Such emission is possible if on average each SAQD contains three or more e-h pairs. Therefore, in these cases the number of peaks and the spectrum shape strongly depend on the intensity of the exciting laser light, which determines the population of carriers within the SAQDs.[10] The spectra reported here, by contrast, do not change their shape for four orders of magnitude variations in the exciting laser intensity. We therefore rule out this possibility.

The evolution of the PL spectrum with InAs deposition time is explained as follows. When the nominal deposited InAs thickness is only 1.2 ML, very few SAQDs are formed. The average distance between QDs is therefore large enough for recombination to occur within the WL. With increasing nominal InAs thickness, the SAQDs density significantly increases. As a result, the emission from the WL drops rapidly, while the PL from the SAQDs increases. In addition, the SAQDs height distribution is also affected by the total amount of InAs deposited. The thicker the InAs thickness is, the higher the average height of the SAQDs is. As a result, the relative intensity of the PL peaks changes, and the lower energy PL peaks, due to e-h recombination within higher SAQDs, become stronger. Thus, the whole spectrum seems to be shifting towards lower energies. In addition, each of the peaks is slightly red shifted, possibly due to the increase in the lateral dimensions of the SAQDs with the increase in the total amount of InAs deposited.

In the second series of samples we kept the nominal thickness of the InAs constant, and varied the composition of the underlying GaInP layer. The PL spectra of this series of samples are shown in Fig. 2. The PL peaks due to the discrete monolayer steps of the SAQDs height are observed for all the dots grown on $Ga_xIn_{1-x}P$ layers. The emission from SAQDs grown directly on InP shows only faint traces of these peaks. Moreover, the PL peaks from this sample are rather wide, and most of the emission is from SAQDs that are more than 5 monolayers high. Similar observations were previously reported for InAs SAQD grown on InP.[4,6] With the increase in the growth interruption time, these peaks disappeared and a broad emission peak at a longer wavelength evolved.[4]

The addition of Ga to the underlying layer drastically changes the size distribution of the SAQDs and enhances the appearance of the distinct peaks in the PL spectrum. With increasing Ga content the concentration of the smaller SAQDs increases, while the concentration of the larger SAQDs decreases. In addition, the various PL peaks shift to higher energies with increasing Ga concentration. The unique advantage of the discrete nature of the PL lineshape is that each peak can be associated with a sub population of SAQDs whose thickness is exactly an integer number of monolayers. This property enables us to study the shift of the emission energies of SAQDs having exactly the same height. Since the height of the SAQDs is unambiguously determined, there are only two possible reasons for the observed blue shift: the larger energy barrier beneath the SAQDs, or the diffusion of Ga atoms from the GaInP layer into the InAs SAQDs.

In order to distinguish between these two explanations we used an eight band, k·P model for calculating the energies of the strained InAs SAQDs.[11] The effect of the strain due to the lattice mismatch between the different layers is inherent in our calculations. Based on the AFM measurements, we could safely assume that the lateral dimensions of the SAQDs are much larger than their height (which is only few monolayers). Therefore, the single carrier energy levels in a SAQD must closely correspond to those of a quantum well (QW) having the same thickness (number of monolayers) as the SAQD height.

In Fig. 3, we compare between the 1.8K measured SAQD PL peak energies and the calculated optical transition energies of QWs of same thickness. The calculations for an InAs QW (i.e. assuming no Ga diffusion into the SAQDs), are in excellent agreement with the measured PL peak positions of the samples with $Ga_xIn_{1-x}P$ layer with $x \leq 0.09$. This agreement also suggests that the As/P exchange reaction at the InAs/InP interfaces, reported to affect mainly the energy levels of the smaller SAQDs,[6,12] can be neglected in our samples.

For the samples with a higher Ga content, however, the agreement is poor. The deviation is larger for the taller SAQDs where the InAs QW calculated energies are significantly lower than the measured ones. We also note that the PL energies of the higher dots are almost independent of the Ga content in the $Ga_xIn_{1-x}P$ layer, since in thick layers the confined levels are deep, and their energies are not sensitive to the height of the barriers.  Therefore, one must assume Ga incorporation within SAQDs.

For a quantitative estimation of the effect of the Ga diffusion into the SAQDs we repeat the calculations for $Ga_yIn_{1-y}As$ QWs, with $y < x$. The Ga content (y) in the $Ga_yIn_{1-y}As$ QW in these calculations was varied in order to best fit the experimentally measured spectra. The results of the calculations are represented in Fig. 3 by lines for the samples with $Ga_xIn_{1-x}P$ with $x = 0.19$ and $x = 0.30$. Excellent agreement is obtained between the calculations and the measured energies of the SAQDs PL peaks.

The fitted Ga content (y) in the SAQDs is presented in the inset to Fig. 3 as a function of the Ga content in the underlying GaInP layer (x). Our results establish that Ga atoms diffuse into the SAQDs during the growth. We note that diffusion of Ga from the underlying layer to the SAQDs decreases the strain in both the layer and the SAQDs. Similar phenomenon of interdiffusion between SAQDs and their surrounding host material was already reported for InAs/GaAs SAQDs, where Ga concentrations in the SAQDs as high as 50% were observed.[13]

In conclusion, extremely thin self-assembled InAs quantum dots of few monolayer thickness, were grown on $Ga_xIn_{1-x}P$ layers. The size distribution of these QDs depends both on the total amount of InAs deposited, as well as on the Ga content in the $Ga_xIn_{1-x}P$ layers. With increasing Ga content in the $Ga_xIn_{1-x}P$ layer the average height of the QDs decreases. Diffusion of Ga atoms into the SAQDs is observed and quantified by means of optical spectroscopy.

**Figure Captions:**

FIG. 1: 77K PL spectra of SAQDs grown on a $Ga_{0.19}In_{0.81}P$ thin layer for InAs nominal thickness ranging from 1.2 to 4.2 ML. The various PL lines are marked in the figure.

FIG. 2: 77K PL spectra of nominally 2.4 ML InAs SAQDs layer on a $Ga_xIn_{1-x}P$ layer for Ga contents ranging from x = 0 to x = 0.3.

FIG. 3: Measured (symbols) and calculated (lines) SAQD PL emission energies at T=1.8K. The inset shows the fitted Ga content in the SAQDs (y) as a function of the Ga content in the GaInP layer (x).

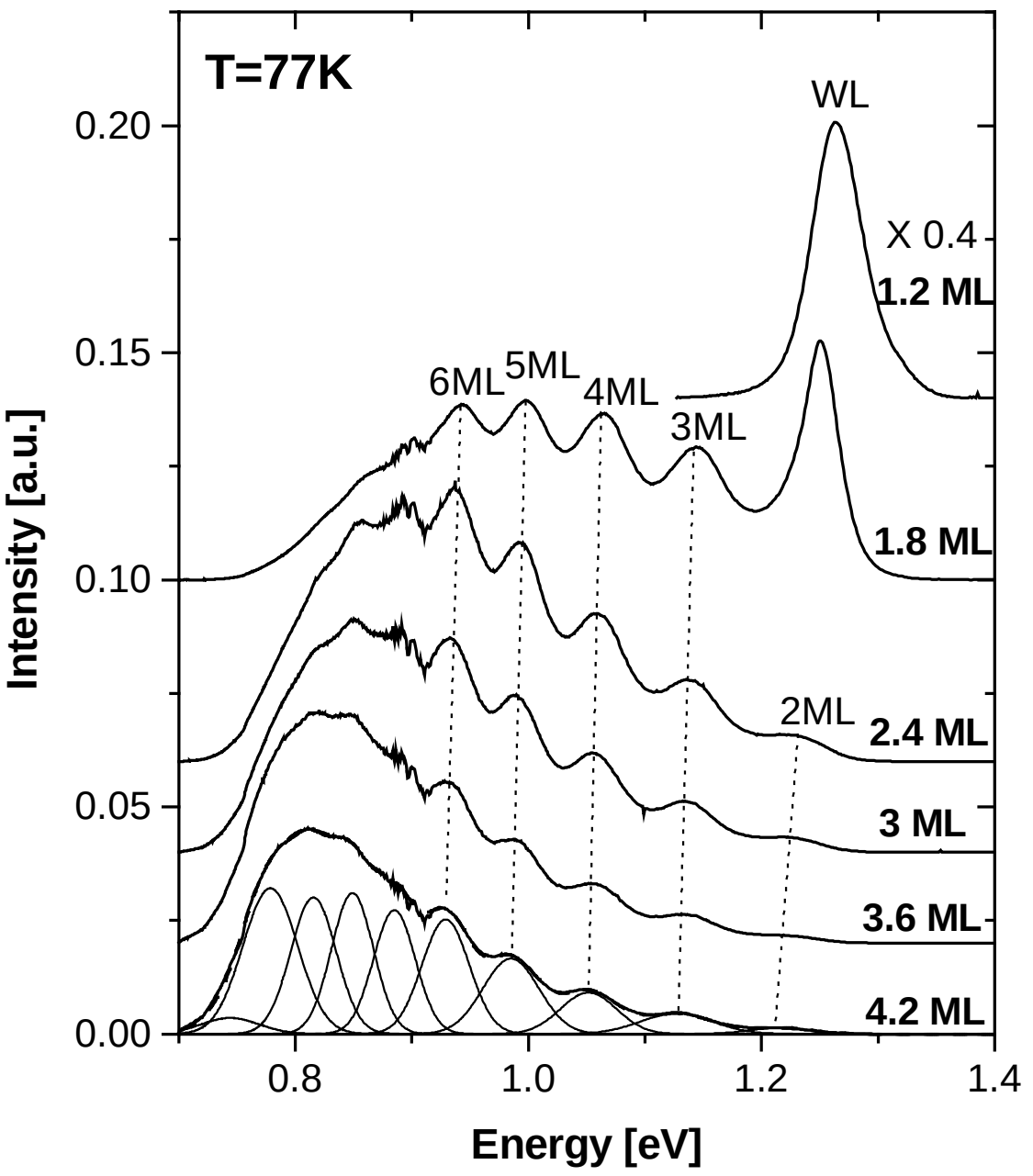

FIG. 1

T. Raz et al.

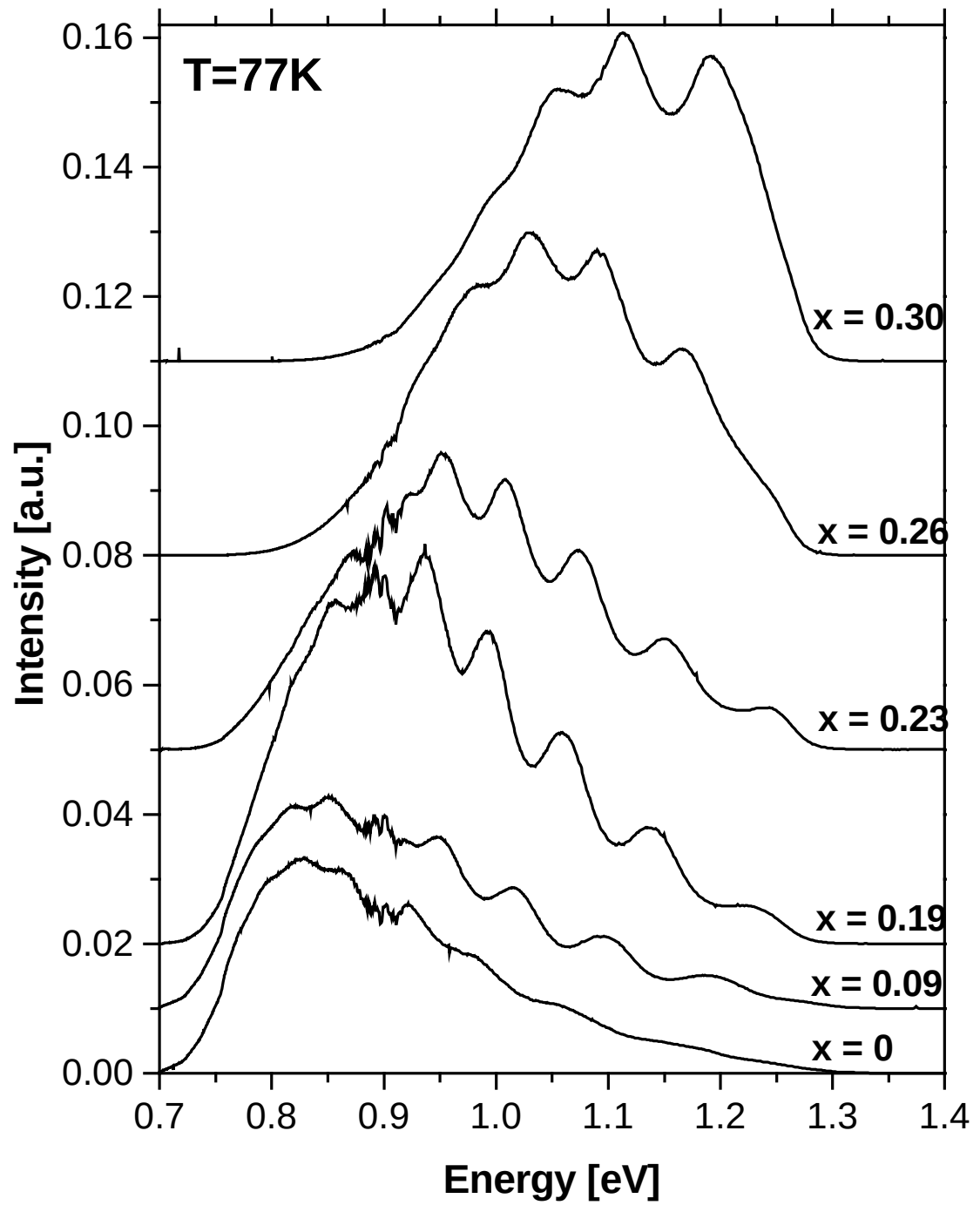

FIG. 2

T. Raz et al.

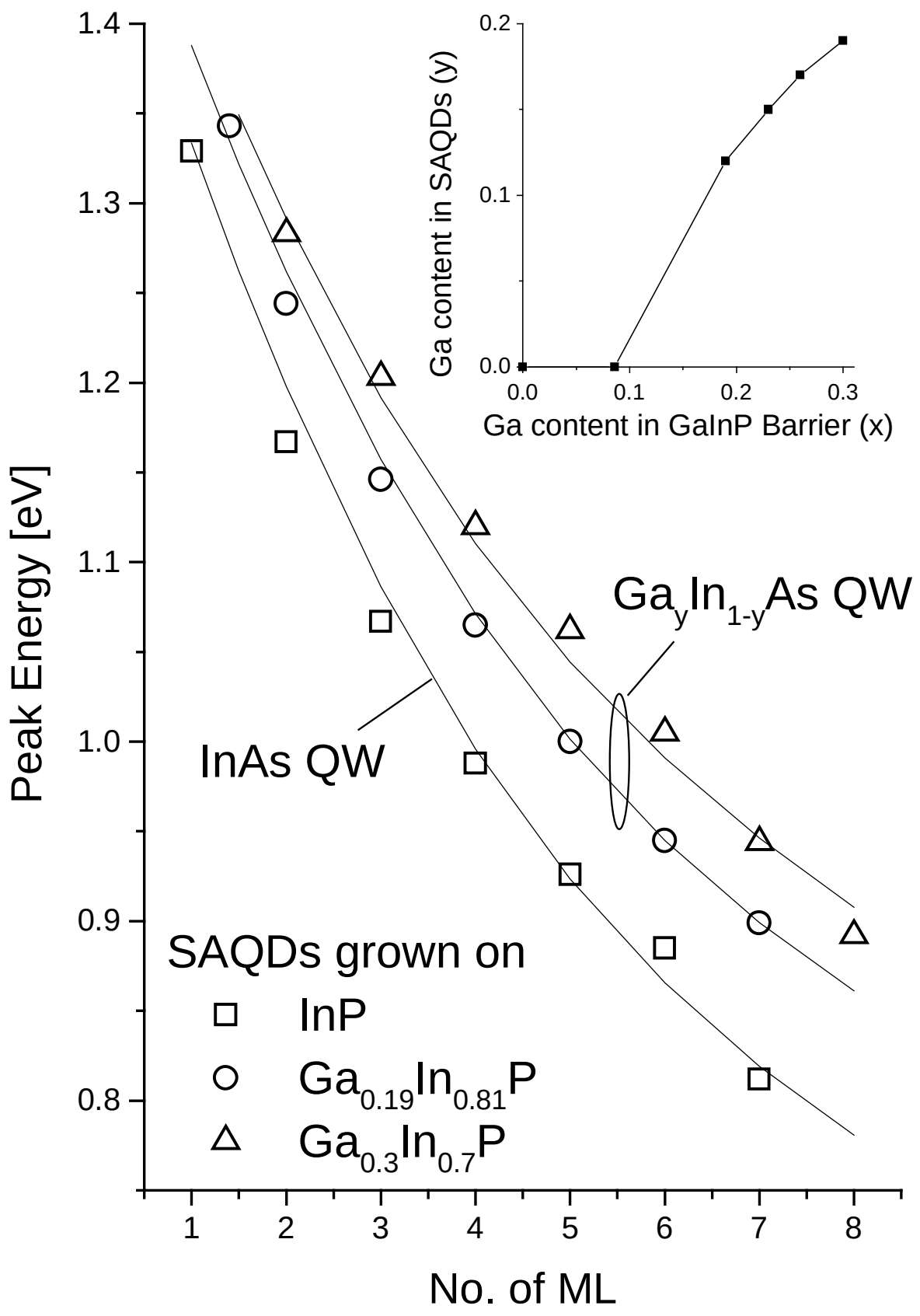

FIG. 3

T. Raz et al.